# Towards Query Optimization for SPARQL Property Paths

Nikolay Yakovets   Parke Godfrey   Jarek Gryz
York University
Toronto, Canada
{hush, godfrey, jarek}@cse.yorku.ca

## ABSTRACT

The extension of SPARQL in version 1.1 with *property paths* offers a type of *regular path query* for RDF graph databases. Such queries are difficult to optimize and evaluate efficiently, however. We have embarked on a project, WAVEGUIDE, to build a cost-based optimizer for SPARQL queries with property paths. WAVEGUIDE builds a query plan—a *waveguide plan* (WGP)—which *guides* the query evaluation. There are numerous choices in the construction of a plan, and a number of optimization methods, meaning the space of plans for a query can be quite large. Execution costs of plans for the same query can vary by orders of magnitude. We illustrate the types of optimizations this approach affords and the performance gains that can be obtained. A WGP's costs can be estimated, which opens the way to cost-based optimization.

## 1. INTRODUCTION

Graph data is becoming rapidly prevalent with the rise of the Semantic Web, social networks, and data-driven exploration in life sciences. There is need for natural and efficient ways to query over these graphs.

The Resource Description Framework (RDF) [19] provides a data model for graph data. An RDF *store* is a set of *triples* that describes a directed, edge-labeled multigraph. A triple, $\langle s, r, o \rangle$, denotes an *edge* from *node* "s" (the *subject*) to node "o" (the *object*), with the edge labeled by "r" (the *role*, also called as the *label* or as the *predicate*).[1]

Correspondingly, the SPARQL query language [18] provides a formal means to query over RDF stores. A query defines sub-graph match criteria; its evaluation over an RDF store returns all embedded sub-graphs meeting the criteria. For example, the query "?friend :friendOf Charles" evaluates to a list people (nodes, binding to variable "?friend") who are friends of (role ":friendOf") "Charles" (a named node, so a constant). This is a simple query, of course, and could be evaluated just by extracting the triples with "r = :friendOf" and "o = Charles". For even an only slightly more complicated query, however, it may not be straightforward to find a *plan* to evaluate it efficiently.

In its current version, 1.1, SPARQL's expressiveness has been extended with *property paths* [11]. Instead of specifying the path of interest *explicitly* between nodes, one may now specify it *implicitly* via a *regular expression*. (This also means matching paths in the graph are not bounded in length by the query's expression, while they are in SPARQL 1.0). For example, the query "?friend :friendOf+ Charles" evaluates to a list people who are friends of "Charles", or friends of people who are friends of "Charles", and so forth.[2]

Property paths effectively introduce the concept of *regular path queries* (RPQs)—well studied before the advent of RDF and SPARQL—into the query language. While eminently useful, such queries are even more challenging to optimize well. We have embarked on a long-term project called WAVEGUIDE with the ultimate goal to provide viable cost-based query optimization and evaluation for SPARQL over RDF stores that is on par with the state of the art for relational database systems.

We address the critical first step of this endeavor, defining a rich *plan space*—the space of *waveguide plans* (WGPs)—for SPARQL queries. We focus on single-path, property-path queries, essentially the RPQ fragment of SPARQL 1.1. We consider a set semantics—the distinct directive in each query—and thus do not consider aggregation. Contributions of this work are as follows.

1. *Waveguide-plan space.*
   (a) *Summarize* the state of the art for evaluation of RPQs and SPARQL property paths (§2). *Establish* why none suffices (§2.4).
   (b) *Devise* the waveguide place space (§3). *Demonstrate* it *subsumes* the state of the art, and extends well beyond it (§3.5).
   (c) *Model* the *cost factors* that determine the efficiency of plans (§4). *Present* the powerful optimizations offered by waveguide plans (§4.3).
2. *Performance study.*
   (a) *Provide* an evaluation framework (§5.1).
   (b) *Benchmark* query plans for realistic queries over real RDF stores / graphs (§5). *Substantiate* the optimizations of our approach (§5.3, §5.4, & §5.5).
   (c) *Justify* the *necessity* of planning and the waveguide plan space (§5).

---

[1] The object in an RDF triple is allowed to be a *literal* as well as a node. However, this distinction is not important for us.

[2] ":friendOf+" represents the *transitive closure* over edges labeled as ":friendOf".

A *waveguide plan* consists of a collection of (non-deterministic) finite automata for the property path and search directives which *guides* the query evaluation. In [22], we demonstrated that, with proper choice of plan, we can gain orders of magnitude performance improvement for many property-path queries over real datasets, while maintaining comparable performance for other queries, as the leading SPARQL query engines as JENA [12] and VIRTUOSO [9]. We evince that planning is critical to evaluate SPARQL queries efficiently, and that choosing the right plan depends on the underlying graph data and thus ultimately must be cost-based.

## 2. BACKGROUND & RELATED WORK

In §2.1, we provide relevant background on path queries. The literature on path queries over graphs, as is pertinent to property paths, comes from two distinct sources:
1. work on *regular path queries* (RPQs); and
2. work on SPARQL platforms to extend to version 1.1 to handle *property paths*.

Research on RPQs, which well precedes RDF and SPARQL, mostly focused on theoretical aspects, but little on performance issues for evaluating such queries in practice.[3] The seminal work that introduced the G+ query language [16] exploited the natural observation that where there is a regular expression, there is a *finite automaton* (FA) that is a *recognizer* for it. They showed how to use finite state machines to direct search over the graph to evaluate a RPQ. In essence, an FA corresponding to the query's regular expression provides a plan for its evaluation. Subsequent work on RPQs followed on this idea. Let us call this the FA approach. We overview this approach in §2.2.

Work on evaluating property paths—much newer by virtue of the fact that the SPARQL 1.1 standard is quite recent—meanwhile has mirrored the dynamic-programming approach behind the algorithm presented in the seminal work of [15]. This can be modeled by an extended *relational algebra* (RA) that includes an operator $\alpha$ for *transitive closure* ($\alpha$-RA) [3]. Let us call this the $\alpha$-RA approach.

As with the FA approach for RPQs, $\alpha$-RA suffices for evaluating property paths. The full power of relational algebra, as extended with $\alpha$, can then be employed to devise an evaluation plan—an $\alpha$-RA-expression tree—based on the regular expression of the property path. This general approach is found behind many SPARQL platforms, as it follows relational techniques well. For example, VIRTUOSO [9], a leading SPARQL system which is also a well-established relational database system, extended their platform to accommodate property paths by adding an "$\alpha$" operator to the engine. We present this approach and characterize it by $\alpha$-RA in §2.3.

Work on property-path evaluation has been remiss in not drawing the connection to RPQs. How do the FA and $\alpha$-RA approaches compare? Does one subsume the other? Or are they *incomparable*? The latter is, in fact, the case, and we show this in §2.4. Furthermore, a combined approach might be superior. We show that it is in §3.

Both the FA and $\alpha$-RA approaches effectively provide evaluation plans for property-path queries. However, the *plan*

---
[3] Regular-path queries have been considered under both *simple*- and *arbitrary*-path semantics. Under simple-path semantics, a path in the graph to match must not repeat any nodes; under arbitrary-path semantics, they may. SPARQL adopts *arbitrary*-path semantics, for the sake of tractability.

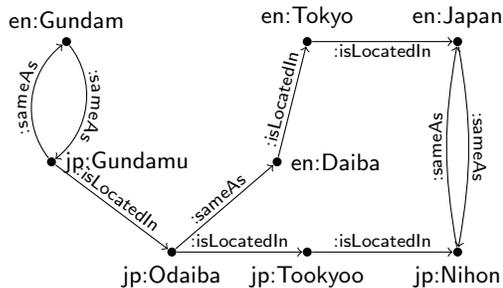

Figure 1: An example graph database.

*spaces* that are implicit in these approaches have not been considered. In FA, choosing a different (but still correct) automaton for the plan might offer a significantly more efficient plan. In systems taking the $\alpha$-RA approach, limited planning is sometimes done, but not in a formal way to enumerate through the plan space to find a best estimated plan, as is done in relational systems. We address this in §4.

### 2.1 Path Queries on Graphs

A *graph database* $G$ can be defined as $\langle N, \Sigma, E \rangle$ for which $N$ is a finite set of nodes (vertices), $\Sigma$ is a finite *alphabet* (a set of *labels*), and $E$ is a set of *directed, labeled edges*, $E \subseteq N \times \Sigma \times N$.

A *path* in a graph is defined as a sequence $p = n_0 a_0 n_1 \ldots n_{k-1} a_{k-1} n_k$ such that $n_i \in N$, for $0 \le i \le k$, and $\langle n_i, a_i, n_{i+1} \rangle \in E$, for $0 \le i < k$. The path-induced *path label* $\lambda(p)$ is the string $a_1 a_2 \ldots a_k \in \Sigma^*$ (for which $\Sigma^*$ is a set of all finite strings formed over $\Sigma$). Each node $n \in N$ is associated with an *empty path*, $n$, the path label of which is the empty string, denoted by $\epsilon$.

A *regular expression* over alphabet $\Sigma$ is defined inductively, as follows: 1. the empty string $\epsilon$ and each symbol $r \in \Sigma$; and, 2. given regular expressions $r$, $r_1$, and $r_2$, then (a) the concatenation $r_1 r_2$, (b) the disjunction $r_1 | r_2$, and (c) Kleene star $r*$. The *regular language* defined by the regular expression $r$ is denoted by $L(r)$. Regular language is defined inductively, as follows: 1. $L(\epsilon) = \{\epsilon\}$ and $L(a) = \{a\}$, for each $a \in \Sigma$; and, 2. for inductively combining strings, (a) $L(r_1 r_2) = L(r_1) \cdot L(r_2)$, (b) $L(r_1 | r_2) = L(r_1) \cup L(r_2)$, and (c) $L(r*) = \{\epsilon\} \cup \bigcup_{i=1}^{\infty} L(r)^i$.

A *regular path query* $Q$ is a tuple $\langle x, r, y \rangle$ for which $x$ and $y$ are free variables (that range over nodes) and $r$ is a regular expression. An *answer* to $Q$ over graph $G = \langle N, \Sigma, E \rangle$ is a pair $\langle s, t \rangle \in N \times N$ such that there exists an arbitrary path $p$ from node $s$ to node $t$ for which the path label $\lambda(p)$ is in language $L(r)$ ($\lambda(p) \in L(r)$). The *answer set* of $Q$ over graph $G$ is the set of all answers of $Q$ over $G$.

Regular path queries have been considered since *semi-structured* data models were first introduced [2, 16]. The complexity of RPQs for graph databases particularly has been well studied [4, 5]. In [14], the idea of employing NFAs to guide search for RPQ evaluation appears. In [10], they perform a fixpoint evaluation for property paths. In [21], we present a precursor of WAVEGUIDE that explores fixpoint evaluation for property paths using SQL recursion.

Regular path queries provide a useful mechanism for querying data in many application domains. For example, consider the knowledge base dataset of the Linked Open Data (LOD) cloud. LOD is a community effort which aims to interlink the structural information available in various datasets on the Web (such as Wikipedia, WordNet, and others), and make it available as a single RDF graph.

RPQs prove useful in querying such linked data by providing a convenient declarative mechanism which can be used to answer queries without prior knowledge of the underlying data paths.

EXAMPLE 1. Consider the part of a LOD graph database as presented in Fig. 1. This represents information the Gundam robot statue in Odaiba in Tokyo. The data has been integrated from two datasets, identified by the prefixes en and jp, standing for the English and Japanese Wikipedia, respectfully. The data entities between these two datasets are interlinked by using OWL ontology terms. Equivalent entities are connected with owl:sameAs edges. In this case, the Japanese dataset contains richer spatial information related to the statue than does the English dataset.

Say a user wants to know in which country this Gundam statue is located. Since there are no direct :isLocatedIn edges outgoing from en:Gundam—as is often the case in linked data—the graph needs to be *searched*. During the search, equivalent data entities need to be resolved by following :sameAs edges. Likewise, a spatial hierarchy needs to be computed by following :isLocatedIn edges. This search can be defined by the following SPARQL query pattern:

$$\text{en:Gundam (:sameAs*/:isLocatedIn)+} \qquad (\mathcal{Q}_1)$$
$$\text{/:sameAs* ?place .}$$

$\mathcal{Q}_1$ computes the spatial hierarchy starting from node en:Gundam, using information from both interlinked datasets to resolve equivalent entity closures.

## 2.2 FA Plans

Regular expressions are a formal notation for patterns that *generate* strings—called *words*—over an *alphabet*. The set of words that a given regular expression can generate is called its *language*.

The dual to generation is *recognition*. *Finite state automata* are the recognition counterpart to regular expressions. For any given regular expression, a finite state automaton—abbreviated as *finite automaton*—can be constructed that will *recognize* the words over the alphabet that belong to the expression's language.

Thus, an FA $A$ can be constructed to recognize the language of a given regular expression $r$. One can construct one such FA by traversing the parse tree of $r$ bottom up, and combining the automata that recognize sub-expressions of $r$ into a composite automaton via union, concatenation, and closure of the sub-automata as is appropriate.

EXAMPLE 2. Recall query $\mathcal{Q}_1$ from Ex. 1. As shown in Fig. 2, an automaton construction for this query is a two-step procedure. First, traversing the parse-tree of $r$ bottom up, the $\epsilon$-NFA is built up, by the base case and the inductive rules. Second, the resulting $\epsilon$-NFA is then *minimized* to an NFA, which typically has smaller size, and hence, is more efficient to process.

The first algorithm to use automata to evaluate regular expressions on graphs was presented in [16] as a part of an implementation of the G+ query language. Given a graph database $G = (N, E)$ and a query $Q = (s, L(r), t)$ in which $s$ and $t$ are nodes in $G$, the algorithm proceeds as follows. The expression $r$ is converted into a finite automaton $A_Q$ by using the bottom up traversal of parse tree of $r$, as discussed. Then, the graph database $G$ is converted to finite automaton $A_G$ with graph nodes becoming automaton states and graph edges becoming transitions. Node $x$ is assigned to be the initial state, and $y$ is assigned to be the accepting state in $A_G$.

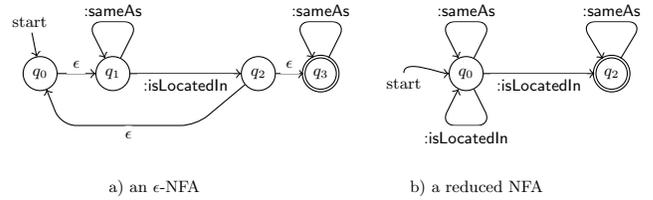

a) an $\epsilon$-NFA      b) a reduced NFA

Figure 2: An $\epsilon$-NFA and corres. reduced NFA for $\mathcal{Q}_1$.

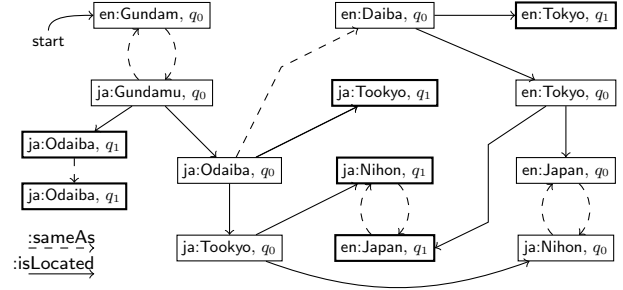

Figure 3: Example product construction of automata.

Then, given $A_G$ and $A_Q$, a product automaton $P = A_G \times A_Q$ is constructed. $P$ is then tested for non-emptiness, which checks whether any accepting state can be reached from the initial state. If the language defined by $P$ is not empty, then the answer for the reachability query $(s, L(r), t)$ on graph $G$ is "yes"; i.e., there exists a path between $s$ and $t$ in $G$ that conforms to $r$. This idea of employing a product automaton for RPQ evaluation over graphs has been used in [6, 13, 14, 16, 17, 23].

EXAMPLE 3. Given query $\mathcal{Q}_1$ and the database $G$ from Ex. 1, the corresponding product automaton $P = A_G \times A_Q$ is shown in Fig. 3. $P$ is a representation of the *search space* that needs to be explored to answer $\mathcal{Q}_1$. $P$ can be explored using any search strategy—e.g., breadth-first search—starting from the initial state $\langle$en:Gundam, $q_0\rangle$. All reachable accepting states (shown in bold) are the answers to $\mathcal{Q}_1$.

## 2.3 $\alpha$-RA Plans

An alternative approach is to use the $\alpha$-extended relational algebra ($\alpha$-RA) to produce evaluation plans for RPQs. The $\alpha$ operator computes the transitive closure of a relation. Let the graph database be represented as a relation of triples $G(s, p, o)$. Let $T = \pi_{1,3}G$; thus $T$ consists of pairs of nodes $\langle s, o \rangle$ such that the pair is connected by a directed edge im the graph. Then $\alpha$ applied to $T$ computes the *least fixpoint* of the following operation:

$$T^+ = T \cup \pi_{1,3}(T^+ \bowtie_{T^+.o=T.s} T) \qquad (\mathcal{E}_1)$$

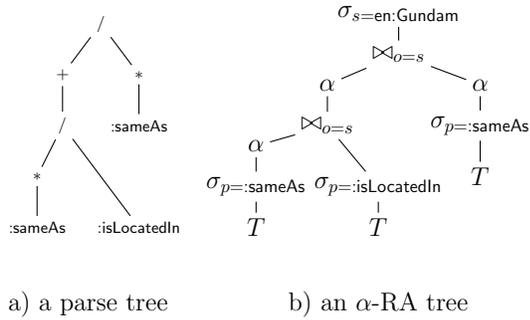

a) a parse tree      b) an $\alpha$-RA tree

Figure 4: A parse tree and $\alpha$-RA tree for query $\mathcal{Q}_1$.

Thus, $\alpha(T)$ results in all pairs of nodes such that, for the nodes of each pair, there *exists* a path between them in the graph (denoted by) $G$. If we were to evaluate the fixpoint by a semi-naïve evaluation, each iteration of evaluation is over paths of length one greater than of the previous iteration. The process stops when no new pairs are added; i.e., the fixpoint has been reached.

Given the SPJRU (select-project-join-rename-union) relational algebra extended with the $\alpha$ operator, one can evaluate the RPQ $Q = (x, L(r), y)$ over graph $G = (N, E)$ by the algorithm proposed in [15]. This traverses the syntax tree of expression $r$ bottom-up. Let $s$ be the sub-expression of $r$ represented by a given node in a parse tree. The binary relation $R_s \subseteq N \times N$ is computed so that node pair $(u, v) \in R_s$ *iff* there exists a path from $u$ to $v$ in $G$ that matches $s$.

The manner in which the relations are joined going bottom-up in a parse tree depends upon the type of the node. The cases are as follows:

1. If $s$ is a $\Sigma$-symbol, then $R_s := \{(u, v) | (u, s, v) \in E\}$.
2. If $s = \epsilon$, then $R_s := \{(u, u) | u \in N\}$.
3. If $s_1$ and $s_2$ are sub-expressions and $s = s_1 | s_2$, then $R_s = R_{s1} \cup R_{s2}$.
4. If $s_1$ and $s_2$ are sub-expressions and $s = s_1 \cdot s_2$, then $R_s = \pi_{1,3}(R_{s1} \bowtie_{R_{s1}.2 = R_{s2}.1} R_{s2})$.
5. If $s = s_1^*$, then $R_s$ is the reflexive and transitive closure of $R_{s1}$, or $R_s = \alpha(R_{s1}) \cup R_{s1}$.
6. If $s = s_1^+$, then $R_s$ is the transitive closure of $R_{s1}$, or $R_s = \alpha(R_{s1})$.

(Correctness of this algorithm is established in [15]).

EXAMPLE 4. Given query $\mathcal{Q}_1$ and the database $G$ from Ex. 1, the corresponding $\alpha$-RA tree is shown in Fig. 4.

The $\alpha$-RA-based RPQ evaluation can be directly implemented in most relational databases and relational triple-stores. In [21], we proposed a method that translates RPQs as defined by SPARQL property paths into recursive SQL. A similar approach was used by Dey et al. [8] in the context of the evaluation of provenance-aware RPQs by a relational engine.

## 2.4 Comparing Plan Spaces

The FA and $\alpha$-RA approaches each entail a *plan space*; that is, the plans collectively an approach produces over all possible property-path queries. Let $\mathcal{P}_{\mathsf{FA}}$ and $\mathcal{P}_{\alpha\text{-RA}}$ denote the plan spaces for FA and $\alpha$-RA, respectively. To understand how the approaches are related—for instance, whether one approach subsumes the other, or whether they are incomparable—we consider these plan spaces. The Venn diagram of how they are related is shown in Fig. 5c.[4]

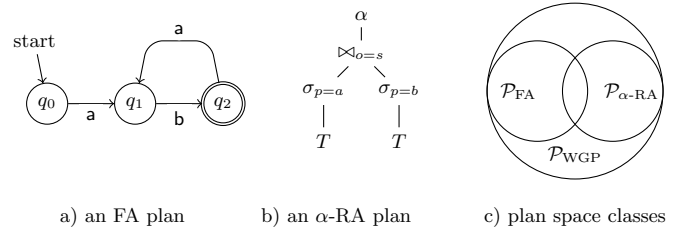

a) an FA plan    b) an $\alpha$-RA plan    c) plan space classes

Figure 5: Example plans.

CLAIM 1. $\mathcal{P}_{\mathsf{FA}}$ and $\mathcal{P}_{\alpha\text{-RA}}$ are *incomparable* ($\mathcal{P}_{\mathsf{FA}} - \mathcal{P}_{\alpha\text{-RA}} \neq \emptyset$ and $\mathcal{P}_{\alpha\text{-RA}} - \mathcal{P}_{\mathsf{FA}} \neq \emptyset$), but *overlap* ($\mathcal{P}_{\mathsf{FA}} \cap \mathcal{P}_{\alpha\text{-RA}} \neq \emptyset$).

Of course, we are taking liberties; the place spaces should be over the same *domain* of plans. As we have presented things, however, they are not; we have presented FA plans as automata and $\alpha$-RA plans as algebraic trees. To prove formally the claim in Fig. 5c, we would need to establish an isomorphism between FA and $\alpha$-RA plans, or have a canonical form for plans to which each plan type could be mapped. This can be done. The formalism for waveguide plans we will present in §3 would suffice for this mapping. DATALOG, or the *relational algebra* extended by *while loops* (established to be expressively equivalent to DATALOG) [1], would provide an even more universal domain that would suffice.

This is beyond the scope of what we can do here. Still, we easily can establish informally that these spaces are distinct. Consider the following *generic* property-path query pattern:

$$\text{?x (a/b)+ ?y} . \qquad (\mathcal{Q}_2)$$

We shall be using $\mathcal{Q}_2$ as a prevalent example. Here, "a" and "b" are stand-ins for labels. It matches node-pairs that are connected by some path labeled ab, abab, or ababab, and so forth. This is a quite simple property-path query, but one that already demonstrates the complexities of planning.

The FA plan in Fig. 5a would be in $\mathcal{P}_{\mathsf{FA}}$ for $\mathcal{Q}_2$. There is no $\alpha$-RA plan that could be equivalent to it, however; none would ever evaluate aba, ababa, and so forth as state $q_1$ does in the FA plan. $\alpha$-RA plans cannot compute transitive closure in a *pipelined* fashion as the FA plan is doing; the $\alpha$ operator acts over an entire relation.

The $\alpha$-RA plan in Fig. 5b would be in $\mathcal{P}_{\alpha\text{-RA}}$ for $\mathcal{Q}_2$. There is no FA plan that could be equivalent to it, however; no state transition in its automata can represent the "join" with ab. FA plans do not encompass *views*, materialized parts of the query that can be reused, while the $\alpha$-RA plan does by effectively materializing ab to join repeatedly on it.

Meanwhile, there are many plans in common between FA and $\alpha$-RA: for any query that is restricted to transitive closure over single labels, for example, will result in common FA and $\alpha$-RA plans.

## 3. WAVEGUIDE PLANS

---
[4]The diagram's claim that the plan space of waveguide plans, $\mathcal{P}_{\mathsf{WGP}}$, properly subsumes both $\mathcal{P}_{\mathsf{FA}}$ and $\mathcal{P}_{\alpha\text{-RA}}$ is taken up in §3.5.

WAVEGUIDE's evaluation strategy is based on an *iterative search algorithm*—and variations of it—which is guided by the WGPs. We are able to express complex query evaluation plans which involve multiple search *wavefronts* that iteratively explore the graph. The *states* of the wavefront automata in a WGP represent path queries in their own right. As the WGP (selectively) materializes states during evaluation—which we call *path views*—this allows wavefronts to re-use intermediate results (paths) that were already discovered by the search process.

### 3.1 Wavefronts

In WAVEGUIDE, we propose a novel strategy to perform efficiently path search while *simultaneously* recognizing the path expressions. WAVEGUIDE's input is a graph database $G$ and a *waveguide plan* (WGP) $P_Q$ which *guides* a number of search wavefronts that explore the given graph. This graph exploration, driven by an iterative search procedure, is inspired by the semi-naïve bottom-up strategy used in evaluation of linear recursive expressions based on *fixpoint*, as is done for the $\alpha$ operator for $\alpha$-RA, described in §2.3.

The key idea is, given a *seed* as a start, to expand repeatedly the search wavefronts in the graph until no new tuples are produced; i.e., we reach fixpoint. Each search wavefront is guided by a *wavefront automaton*, a finite state machine based on non-deterministic finite automata (NFA). This is akin to the FA approach discussed in §2.2. Different, though, from NFAs which are used as recognizers of regular expressions on strings, wavefront automata introduce a number of features directed to evaluation of regular expressions on graphs. These include the use of *seeds*, *append* and *prepend transitions*, and *path views*.

First, we present the iterative procedure used in WAVEGUIDE that drives the wavefront expansion. Next, we describe the new types of transitions enabled by the wavefront data-structure. Finally, we discuss the interactions between different wavefronts guided by a plan, which can be used for optimization.

### 3.2 Expanding a Wavefront

Each search wavefront has a *seed* as its initialization. The seed is the set of nodes in the graph from which this wavefront begins its search. A seed can be either *universal* or *restricted*. A wavefront with a universal seed conducts its search effectively starting from *every* node in a graph. A wavefront with a restricted seed is restricted to starting search just from those nodes in its seed. (A restricted seed will be defined by the results of other wavefronts or by constants used in a query.) Graphically, a seed is represented as an incoming edge to starting state $q_0$ of the wavefront. We use the label "$U$" to denote a universal seed; any other label on this edge denotes a restriction placed on the seed, thus a restricted seed.

Given an evaluation plan defined by search wavefronts, the graph exploration is performed by an iterative procedure as illustrated in Fig. 6. For example, consider WGP $P_1$ that uses a *single* search wavefront to answer query $Q = (x, (ab)+, y)$ on graph $G$ as shown in Fig. 7. Let the wavefront $W_1$ be constructed by a direct mapping of the query's regular expression into an NFA.

During the search, intermediate results are kept in a *cache*, denoted at iteration $i$ by $C_i$. This is a collection of tuples $\langle u, v, s \rangle$ for which $u$ and $v$ are nodes in $G$ and $s$ is a state

```
WAVEGUIDESEARCH(G, A_Q)

1      Δ_0^R ← seed(G);
2      i ← 0;
3      while |Δ_i^R| ≥ 0 do
4          Δ_{i+1}^S ← seed(Δ_i^R);
5          Δ_{i+1}^C ← crank(Δ_{i+1}^S, Δ_i^R, G, C_i, A_Q);
6          Δ_{i+1}^R ← reduce(Δ_{i+1}^C, Δ_i^R, C_i);
7          C_{i+1} ← cache(Δ_{i+1}^R, C_i);
8          i ← i + 1;
9      done;
10     return extract(C_i);
```

Figure 6: WAVEGUIDE evaluation procedure.

in $W_1$. The newly discovered tuples found in the current iteration are denoted by a *delta* $\Delta_i$. We use the cache $C_i$ and the delta $\Delta_i$ to eliminate intermediate answers we have already seen in the search.

In the first step of the search procedure, all the universal seeds are initialized. Specifically, $\Delta_0^R$ is assigned the set of $\langle u, u, q_0 \rangle$ for all $u \in N$;[5] $q_0$ is the starting state for all wavefront automata with universal seeds.

Next, we loop over iterative steps. In each iteration, four operations are performed seed, crank, reduce, and cache. The iteration continues until fixpoint is reached.

The seed step populates the restricted seeds, according to their respective seed conditions. The crank step transitions from the previous delta to the current, $\Delta_i^R \to \Delta_{i+1}^C$. For each node $v$ in $\langle u, v, s \rangle \in \Delta_i^R$, for edge $\langle v, a, w \rangle \in G$ and graph transition $\langle s, a, t \rangle \in W$, $\langle u, w, t \rangle$ is added to $\Delta_{i+1}^C$. Thus crank advances the search *simultaneously* in the graph *and* in the automaton.

To prevent unbounded computation over cyclic graphs, the delta is *reduced*: $\Delta_{i+1}^C$ is checked against both the previous delta $\Delta_i^R$ and the cache $C_i$; tuples that are seen in either $\Delta_i^R$ or $C_i$ are removed to produce $\Delta_{i+1}^R$. Lastly, the cache is updated $C_{i+1}$ by adding the tuples in the reduced delta $\Delta_{i+1}^R$ to it ($C_i$). The iteration halts once $\Delta^R$ is empty.

Recall $W_1$ in this example was produced by directly mapping the regular expression $r$ to an NFA. As the NFA is a recognizer for $r$, it can be established by structural induction that, for any tuple $\langle u, v, s \rangle$ in the cache ($C$) such that $s$ is an accepting state, the pair of nodes $\langle u, v \rangle$ must have a path between them in the graph that conforms to $r$. Thus, WAVEGUIDE produces the correct results. The answer set can be then *extracted* from the cache by selecting the tuples $\langle u, v, s \rangle$ for all accepting states $s$ of automaton $W_1$.

EXAMPLE 5. Consider the wavefront search in Fig. 7 for a query with regular expression $r = (ab)+$ on graph $G$. Plan $P_1$ uses a single wavefront $W_1$ which is a basic wavefront embodying an NFA that recognizes $r$.

For each iteration $i$ of the search, cache $S_i$, delta $\Delta_i$, and reduced delta $\Delta_i^R$ are shown. The search stops when all newly generated tuples are, in fact, duplicates, due to cycles in $G$. The cache tuples that are in accepting state $q_2$ (shown shaded) are then extracted as the answer set.

### 3.3 Guiding a Wavefront

---
[5]This can be optimized to pull just the tuples from the triple store that can participate in the first step of any path to an answer. We call this *first-hop* optimization.

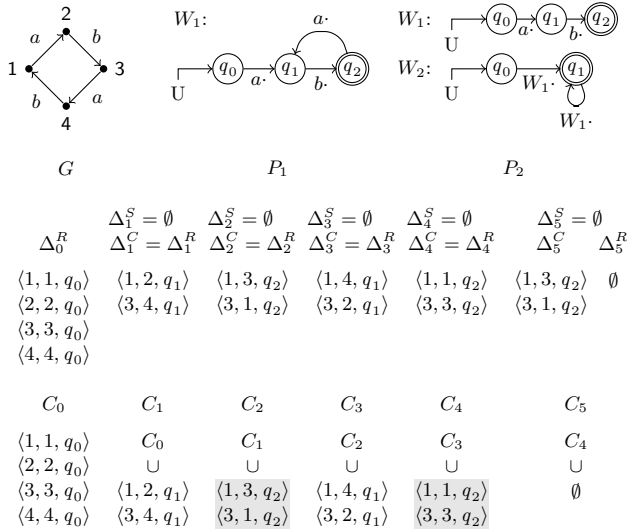

Figure 7: A run of waveguide plan $P_1$ over graph $G$.

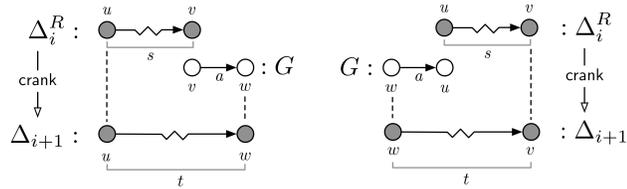

(a) Prepending vs. appending by expanding the wavefront using tuples from graph $G$.

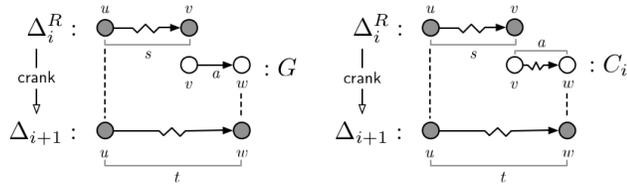

(b) Expanding the wavefront by appending and using the tuples from $G$ (over the graph) vs. from search cache $C$ (over the view).

Figure 8: Types of transitions used in a wavefront.

The construction of the NFA forces an order to the query evaluation. A "wrong" choice of NFA can lead to an inefficient evaluation plan. In WAVEGUIDE, we aim to minimize the search space explored by considering the possible orders of graph exploration by search wavefronts. To achieve this, we use wavefront automata which can use transitions that expand the wavefront in the direction opposite to the direction of the edges of the graph.

Consider a graph transition $\langle s, l, t \rangle$ in wavefront $W$. Edge label $l$ has a general form $\cdot a$ or $a \cdot$, where $a$ is an edge label in $G$. Position of a dot $\cdot$ specifies a direction of a search wavefront and denotes prepend ($\cdot a$) or append ($a \cdot$) transition. Prepend wavefront expands in the opposite direction to the edges in the graph. On the other hand, append parameter guides a wavefront that expands in the same direction as the edges in the graph. The semantics of crank operation on prepend and append transitions are illustrated in Fig. 8a.

Hence, wavefronts enable automaton transitions that explore the graph in a direction specified by the transition. This allows to define a wavefront that can initiate evaluation from any label in the given regular expression and iteratively expand by appending or prepending path labels. This gives us the power to explore all different expansion orders of a single wavefront.

### 3.4 Wavefront Interaction

Often, the search space is constrained even further if several wavefronts are employed in the evaluation, each evaluating parts of a given regular expression. WAVEGUIDE enables this by defining a number of automata, one for each search wavefront.

WAVEGUIDE plans, in addition to transitions over graph edge labels, allow transitions over *path views*, by utizing cached result sets produced by other wavefronts. Consider a transition $\langle s, l, t \rangle$ in $W_1$. If $a$ is an edge label in $G$, then this *graph* transition expands the wavefront by using the tuples from *graph* $G$. Otherwise, if $a$ is a state in $W_2$, then this *view* transition expands the wavefront by employing the tuples produced by wavefront $W_2$ (as illustrated in Fig. 8a).

These new types of transitions offer powerful choices in WGPs for guiding the search. The search can have *multiple* wavefronts originating from different starting points *and* expanding in different directions. Further, each wavefront can employ the cache through transitions over views to avoid unnecessary recomputation.[6]

EXAMPLE 6. Consider the wavefront search in Fig. 7 for a query $Q$ with regular expression $r = (ab)+$. $P_1$ is a basic WGP embodying an NFA that recognizes $r$. From $P_1$, we can design a more efficient WGP, $P_2$: first, compute $(ab)$ with wavefront $W_1$; then use a loop-back view transition to compute the closure $(ab)+$ (with wavefront $W_2$). In this case, it can be shown that $P_2$ explores a smaller search space in fewer iterations than $P_1$.

### 3.5 The WAVEGUIDE Plan Space

We claim that the space of waveguide plans subsumes that of the FA and $\alpha$-RA approaches, as the Venn diagram in Fig. 5 shows (and with the caveats as discussed in §2.4).

CLAIM 2. $\mathcal{P}_{\mathsf{WGP}}$ *properly subsumes* the union of $\mathcal{P}_{\mathsf{FA}}$ and $\mathcal{P}_{\alpha\text{-RA}}$ ($\mathcal{P}_{\mathsf{WGP}} \supsetneq \mathcal{P}_{\mathsf{FA}} \cup \mathcal{P}_{\alpha\text{-RA}}$).

That $\mathcal{P}_{\mathsf{WGP}}$ subsumes each of $\mathcal{P}_{\mathsf{FA}}$ and $\mathcal{P}_{\alpha\text{-RA}}$ is straightforward; we devised WGP so that we could express both FA- and $\alpha$-RA- type plans. WGP extends the FA model. WGP encompasses $\alpha$-RA by the addition of *views*; what the $\alpha$ operator offers, transitive closure over an arbitrary relation, can be accomplished by view-labeled transitions in a waveguide plan.

That $\mathcal{P}_{\mathsf{WGP}}$ properly subsumes the union of $\mathcal{P}_{\mathsf{FA}}$ and $\mathcal{P}_{\alpha\text{-RA}}$ means that there is a waveguide plan that corresponds to no FA plan and to no $\alpha$-RA plan. We have well demonstrated that in the discussions above. Any WGP with multiple wavefronts and some wavefront with a long loop-back is such a plan; FA plans are essentially single wavefront by the FA model, and pipelined loop-backs are outside the scope of $\alpha$-RA. Likewise, any WGP, even single wavefront, that is "mixed", that combines views and long loop-backs, corresponds to no FA plan and to no $\alpha$-RA plan. (In Fig. 12 on page 8, $P_2$ with partial loop-caching is such a plan.) Of course, these very types of waveguide plans that FA and $\alpha$-RA miss often are the most efficient plans for a given query.

---
[6]This is also known as *memoization*.

In §4, we explain why this rich plan space is relevant. In §5, we we compare plans for real queries over real graph data to establish that this is true in practice, as well.

## 4. PLAN COSTS

For a given query, of course, there may be many ways to guide the search. We summarize a *cost framework* for WAVEGUIDE search, *search cost factors* that can magnify the cost (properties of the graph and of resulting pre-paths computed during evaluation), and *optimization methods* that are enabled by WGPs which address the search factors, in turn.

### 4.1 Cost Framework

Recall the three steps in Fig. 6 of the search iteration: crank, reduce, and union. Assume that the search completes in $n$ iterations. The cost of crank, $\mathbf{C}_{\mathsf{crank}}$, corresponds to the total number of *edge walks* performed. This *search size* is the sum of sizes of the deltas. The cost of reduce, $\mathbf{C}_{\mathsf{reduce}}$, has two components: duplicate removal within a delta and for the delta against the search cache. Cost of removal against the delta is often cheaper, since in can be implemented in-memory, while checking against the cache, due to its larger size, might require implementation on secondary storage, therefore increasing its cost.

The cost of union, $\mathbf{C}_{\mathsf{reduce}}$, is associated with search cache maintenance procedures (e.g., indexing), and depends on the size of the search cache.

1. $\mathbf{C}_{\mathsf{crank}} = \sum_{i=0}^{n} f_1(|\Delta_i|)$
2. $\mathbf{C}_{\mathsf{reduce}} = \sum_{i=0}^{n} (f_2(|\Delta_i|) + f_3(|C_i|))$
3. $\mathbf{C}_{\mathsf{union}} = \sum_{i=0}^{n} f_4(|C_i|)$

The cost functions $f_{1-4}$ above are monotone over their parameters; these simply abstract the actual costs as based upon the underlying implementation of WAVEGUIDE's data structures and algorithms.

### 4.2 Search Cost Factors

Properties of the graph and of the WGP chosen—so the guided search as performed in terms of the pre-paths that are computed by the search—will determine the evaluation cost.

*Search Cardinalities.* The *wavefront*, or *wavefronts*, that we choose—as dictated by the wavefronts of the WGP—for the search determines the intermediate results (pairs of nodes connected by valid pre-paths) that we collect each iteration. Just as with different join orders in relational query evaluation, different wavefronts will result in different intermediate delta sizes. These intermediate cardinalities can vary widely from plan to plan.

*Solution Redundancy.* After much deliberation in the research community, the W3C has adopted a *non-counting* semantics for SPARQL property-path queries. Each node pair appears at most once in the answer, even if there are several paths between the node pair satisfying the given regular expression.

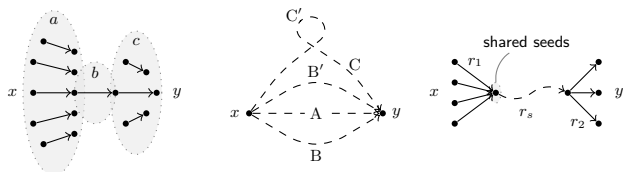

a) search cardinality  b) solution redundancy  c) sub-path sharing

Figure 9: Types of search cost factors.

Answer-path redundancy arises from two sources. First, in dense graphs, solutions are re-discovered by following conforming, yet different paths. Second, nodes are revisited by following cycles in the graph. Thus, the same answer pair may be discovered repeatedly during evaluation. It is critical to detect such duplicate solutions early in order to keep the search size and search cache small.

*Sub-path Redundancy.* In *solution redundancy*, an answer pair could have multiple paths justifying it. Likewise, the paths justifying multiple answer pairs may *share* significant segments (*sub-paths*) in common.

This arises, for instance, in dense graphs and with hierarchical structures (e.g., *isA* and *locatedIn* edge labels). Consider a query "?p :locatedIn+ Canada". Every person located in the neighborhood of the Annex in the city of Toronto qualifies, since the Annex is located in Toronto which is located in Ontario which is located in Canada. The sub-path "Annex :locatedIn+ Canada" is shared by the answer path for each Annex resident.

Because we keep only node-pairs (plus state) in the search deltas, and not explicitly the paths themselves,[7] we may walk these sub-paths many times, recomputing "Annex :locatedIn+ Canada" for each Annex resident.

### 4.3 Plan Optimizations

We consider WGP-optimization methods in relation to the search cost factors above.

*Choice of Wavefronts.* The direction in which we follow edges, and where we start in the graph, with respect to the regular expression will result in different *search cardinalities*. Our choice of automata in the WGP dictates the wavefront(s). For example, consider query $Q = (x, (abc), y)$ and a fragment of a graph shown in Fig. 9a. Since labels $a$, $b$ and $c$ have different cardinalities, different wavefronts will have different search sizes. Consider two plans $P_1$ and $P_2$ that evaluate $Q$ shown in Fig. 10. $P_1$ has a single wavefront that explores the graph starting from $a$, appending $b$ and then $c$. On the other hand, $P_2$ has a wavefront that starts from the low cardinality label $b$, appends $c$ and then prepends $a$. Observe that, in this scenario, $P_2$ results in fewer edge walks than $P_1$.

To reduce overall search size, we need to choose wavefronts that result in fewer edge walks. Wavefronts can be costed to estimate their search sizes based on statistics about the graph, such as 1-gram and 2-gram label frequencies. (Such graph statistics can be computed offline for this purpose.)

*Reduce.* WAVEGUIDE's evaluation strategy is designed to counter solution redundancy. As shown in Fig. 9b, we con-sider several types of redundant solutions based on a path which was followed to obtain each solution. Path A is a shortest path. Paths B and B′ are of the same length, but go through different nodes. Finally, path C′ shares some nodes with path C, but it is longer due to a cycle.

---

[7] Note this design choice in our evaluation strategy is critical for good performance due to solution redundancy!

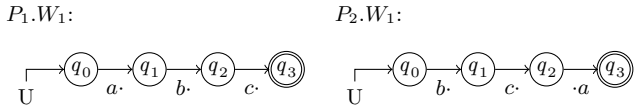

Figure 10: Choosing the wavefronts.

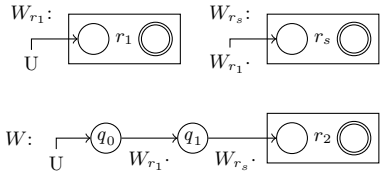

Figure 11: Threading a shared sub-path.

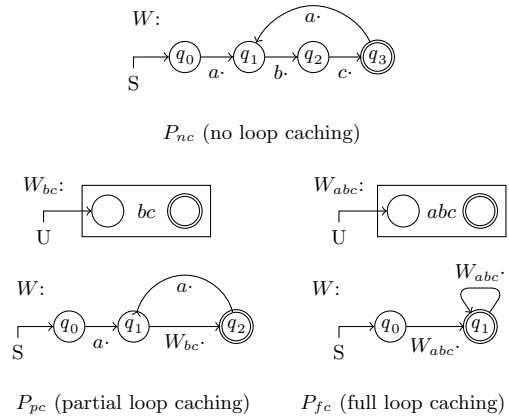

Figure 12: Types of loop caching.

In WAVEGUIDE, redundancy of *candidate* solutions is addressed by removal of duplicates against both cache (cache) and delta (delta) by the reduce operation. Assuming BFS search strategy, duplicate solutions obtained by following paths of the same length (B, B′) are removed within a delta. On the other hand, duplicates obtained by following paths of different lengths (A, B) are removed when delta is compared against a cache. This also includes paths with cycles such as C and C′ as they also have different length.

As a further optimization, once a solution seed-target pair has been discovered, *first-path pruning* (fpp) removes the *seed* from further expansion by the search wavefronts. In our example, once path A has been discovered and solution $(x, y)$ has been obtained, all longer paths (B, B′, C, C′) are never even materialized.

*Threading.* To counter *sub-path redundancy* requires us to decompose a query into sub-queries. We call this decomposition *threading*, and our WGPs accommodate this.

Consider query $Q = (x, (r_1/r_s/r_2), y)$ where sub-path $r_s$ is shared among many solutions as shown in Fig. 9c. This query can be threaded as follows. First, pre-path $r_1$ is computed by wavefront $W_{r_1}$. Then, the portion of the regular expression that will result in sub-paths that will be shared by many answer paths can be computed by a separate wavefront $W_{r_s}$. Here, $W_{r_1}$ seeds wavefront $W_{r_s}$ which computes a shared path for each of the partial solutions produced by $W_{r_1}$. Finally, the complete path is pieced together by wavefront $W$.

Such sub-path sharing can be predicted by graph statistics to indicate when sub-queries should be considered.

*Partial Caching.* Delta results are cached during evaluation as we need to check against the cache for redundantly computed pairs. For large intermediate cardinalities, this can be a significant cost. However, some of this cost can be negated. In particular, not every state in the WGP's automata needs to have its node-pairs cached. Caching is only needed when *unbounded* redundancy is possible, due to cycles in the wavefront automata or in the graph. States without cycles need not be cached.

*Loop Caching.* Transitions over views in wavefront automata allows us to cache and re-use some of the intermediate node pairs we encounter during the search. Such named result sets are useful in reducing unnecessary re-computation by employing an optimization we call *loop caching*.

In transitive query $Q = (x, (r)+, y)$, the expression $r$ is evaluated repeatedly until no new solutions are found. Loop caching rewrites an evaluation plan such that the base $r$ is cached either fully or partially to speed up the transitive evaluation of $(r)+$.

Consider three plans $P_{nc}$, $P_{pc}$ and $P_{fc}$ for query $Q = (x, (abc)+, y)$ shown in Fig. 12. Plan $P_{nc}$ has no loop caching as it evaluates full expression $(abc)$ in a loop. Plan $P_{pc}$ uses a separate wavefront to evaluate $(bc)$ first, then these results are used in a loop to evaluate transitive $(abc)+$. Finally, plan $P_{fc}$ caches full base expression $(abc)$, which is then used in evaluation of a transitive expression.

### 4.4 Cost Analysis

In this section we analyze the cost of plan optimizations that are exclusive to WAVEGUIDE approach, such as threading and loop caching, with relation to the cost model presented in Section 4.1.

*Costs of threading.* Given a plan $P$ with a single wavefront which computes a regular expression of a form $r = r_1/r_s/r_2$, threading rewrites it into a plan $P_t$ with three wavefronts $W_{r_1}$, $W_{r_s}$ and $W_{\text{join}}$ as described in Section 4.2. Regardless of the split of $r$ into $r_1$, $r_s$ and $r_2$, this optimization requires an additional cost of an extra join in $W_{\text{join}}$. If a shared sub-path $r_s$ is accurately identified, then the total reduction of number of edge walks in $P_t$ is sufficiently large to offset the cost of the extra join.

A useful graph metric in identifying the threading split is a *multiplicity* ratio of an expression $r$ in graph $G$, which is computed by analyzing the paths in $G$:

$$\mathcal{M}(G, r) = \frac{|S_s|}{|S_o|},$$

where $S_s$ and $S_o$ is a set of subjects and objects, respectively, connected in $G$ with paths conforming to $r$. Then, $\mathcal{M}(G, r) > 1$, would indicate that, on average, there are many subjects connected to a single object in $G$, while $\mathcal{M}(G, r) < 1$ would indicate that the opposite is true. The greater $\mathcal{M}$ is, the more subjects are connected to the same object, and, hence, more subjects share a path which originates from this object.

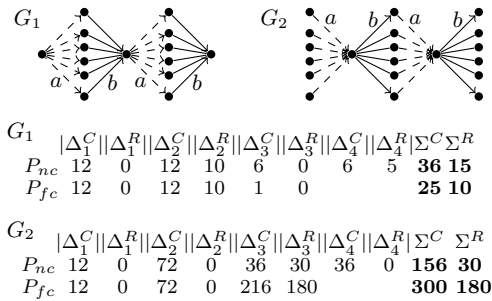

Figure 13: Lensing.

| $G_1$ | $|\Delta_1^C|$ | $|\Delta_1^R|$ | $|\Delta_2^C|$ | $|\Delta_2^R|$ | $|\Delta_3^C|$ | $|\Delta_3^R|$ | $|\Delta_4^C|$ | $|\Delta_4^R|$ | $\Sigma^C$ | $\Sigma^R$ |
|---|---|---|---|---|---|---|---|---|---|---|
| $P_{nc}$ | 12 | 0 | 12 | 10 | 6 | 0 | 6 | 5 | **36** | **15** |
| $P_{fc}$ | 12 | 0 | 12 | 10 | 1 | 0 | | | **25** | **10** |

| $G_2$ | $|\Delta_1^C|$ | $|\Delta_1^R|$ | $|\Delta_2^C|$ | $|\Delta_2^R|$ | $|\Delta_3^C|$ | $|\Delta_3^R|$ | $|\Delta_4^C|$ | $|\Delta_4^R|$ | $\Sigma^C$ | $\Sigma^R$ |
|---|---|---|---|---|---|---|---|---|---|---|
| $P_{nc}$ | 12 | 0 | 72 | 0 | 36 | 30 | 36 | 0 | **156** | **30** |
| $P_{fc}$ | 12 | 0 | 72 | 0 | 216 | 180 | | | **300** | **180** |

Another useful metric is an average length $\mathcal{L}(G, r_s)$ of a path which conforms to $r_s$ in $G$. The longer the shared subpath $r_s$ is, the more potential savings in edge walks can be realized by threading split on $r_s$.

Then, given $\mathcal{M}$ and $\mathcal{L}$ for sub-expressions of $r$ in $G$, the identification of an efficient threading split $r = r_1/r_s/r_2$ becomes an $(\mathcal{M}, \mathcal{L})$ maximization problem.

*Costs of loop caching.* Given a plan with a single wavefront which computes closure $(r)+$ of a regular expression $r$, loop caching rewrites it into a plan in which parts of $r$ are pre-computed, cached, and then used in an iterative evaluation of a closure. For example, consider the differences in evaluation of query $Q = (x, (abc)+, y)$ with plans $P_{nc}$, $P_{pc}$ and $P_{fc}$ shown in Fig. 12. $P_{nc}$ defines a single wavefront, which, due to absence of transitions over views, can be executed pipelined. On the other hand, $P_{pc}$ and $P_{fc}$ first compute $(bc)$ and $(abc)$, respectively, in separate wavefronts, the results of which are used in a wavefront which computes the final closure. Note that due to shorter cycles in wavefront automata in cached plans $P_{pc}$ and $P_{fc}$, the total number of concatenations performed is smaller than in $P_{nc}$. However, the cost of each concatenation is different due to different sizes of the participating relations. For example, $P_{nc}$ concatenates intermediate paths with $a$, $b$ and then $c$, while $P_{fc}$ does the same with a single concatenation with cached $C_{abc}$. In fact, depending on cardinalities of $|C_a|$, $|C_b|$, $|C_c|$, $|C_{bc}|$ $|C_{abc}|$, the concatenations performed in any of the above plans might become the preferred cheaper alternative.

Further, the number of pruned tuples in plans with or without caching can significantly differ depending on the general shape of the graph. For example, consider two basic graphs $G_1$ and $G_2$ as presented in Fig. 13. Both $G_1$ and $G_2$ have the same frequencies of labels $a$ and $b$, but are different in terms of their *shape*. $G_1$ exhibits *lensing* with focal points on concatenations $b/a$, while $G_2$ has lensing in $a/b$. Intermediate cardinalities of $\Delta^C$ (number of edge walks) and $\Delta^R$ (number of pruned tuples) of the WAVEGUIDE search are presented for plans with $(P_{fc})$ and without $(P_{nc})$ loop caching. Observe that loop caching optimization is beneficial for search in $G_1$ with 30% and 33% less edge walks and pruned tuples, respectively. On the other hand, loop caching performs worse in $G_2$ with 92% and 600% more edge walks and pruned tuples, respectively. This can be explained by analyzing the edge walks and pruned tuples during the concatenation sequence $(\ldots/a/b)$ which is performed in $P_{nc}$, but not in $P_{fc}$. In $G_1$, $(\ldots/a/b)$ computes a large number of intermediate tuples most of which are later pruned due to a focal point in $b/a$. Meanwhile, in $G_2$, $(\ldots/a/b)$ first prunes many tuples due to a focal point in $a/b$, hence reducing the total number of edge walks performed later in the search.

Lastly, we consider queries with constants. In pipelined plan $P_{nc}$, this constant can be pushed to seed condition $S$ of its wavefront. In fact, full concatenation $(abc)$ might not need to be ever computed in $P_{nc}$. On the other hand, plans $P_{pp}$ and $P_{fc}$ allow at most partial constant pushdown, since cached relations must be computed with universal seed to ensure completeness of the final closure.

## 5. PERFORMANCE STUDY

### 5.1 The WAVEGUIDE Prototype

We have prototyped a WAVEGUIDE system that implements the methodology from §3 in order to benchmark waveguide plans to study their performance. In this WAVEGUIDE system, resource-intensive tasks are delegated to POSTGRESQL via SQL and *procedural* SQL routines. This implementation of our methodology provides high performance, scalability, and rapid deployment.

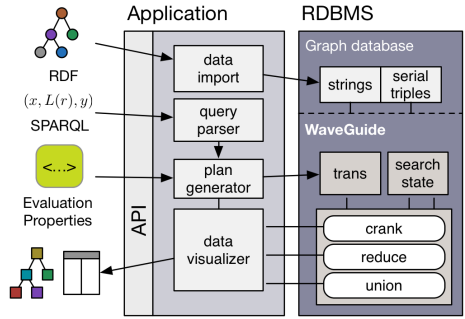

Figure 14: Overview of a prototype system

Fig. 14 shows the architecture. It consists of two layers: *application* and *RDBMS*. The application layer provides a user front-end, preprocessing the graph data, parsing user queries, generating WGPs, and visualizing key steps during the search. The RDBMS layer provides postprocessing of the graph data and performing the iterative WAVEGUIDE graph search for the given WGP.

### 5.2 Methodology

We test our implementation of WAVEGUIDE by running a collection of realistic path queries over real-world datasets YAGO2s [20] and DBPedia [7]. The datasets were preprocessed by removing invalid and duplicate triples and self-loops. After preprocessing, YAGO2s had 242M triples and DBPedia had 463M triples, with 104 and 65K distinct predicates, respectively. This makes these datasets well suited for benchmarking of path queries.

At the time of this paper, we could not find any available benchmarks for SPARQL property-path queries. We therefore generate path queries based on data patterns we identified in real-world graphs. The goal of these experiments is to verify the gains offered by WAVEGUIDE optimizations, and show that they correspond to the cost framework (§4.1) and analysis (§4.4).

Our benchmark was executed on a 2xXeon E5-2640v2 CPU server with 7200RPM HDD running Ubuntu Server 12.04 x64 and PostgreSQL 9.3.

### 5.3 Threading

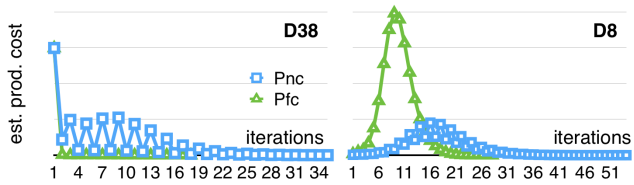

Figure 15: Estimated cost of intermediate concatenations.

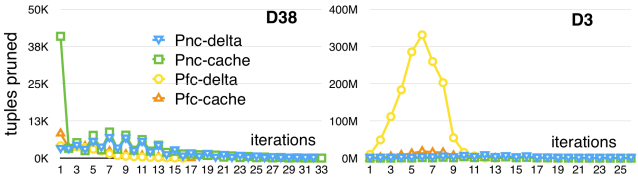

Figure 16: Tuples pruned w/ & w/o loop caching.

We benchmark the threading optimization by executing a query of the following template pattern

$$\text{?x } p/\text{:locatedIn+}/\text{:dealsWith+ ?y} \qquad (\mathcal{Q}_3)$$

over the YAGO2s dataset, with "$p$" as a variable predicate. We chose this template for the following reasons. First, since $\mathcal{Q}_3$ contains the concatenation of two transitive closures, it is difficult to predict the average length of the paths in the answer. Second, locatedIn+ is a popular predicate which also concatenates with many other predicates, so there are many candidates for $p$. Finally, locatedIn+ has an $\mathcal{M}$ value of 11.27 which makes it a good candidate for a threading split in $Q$.

We group $p$ candidates in two sets: the first (queries **L1**–**5**) having an $\mathcal{M}$ value greater than 10; and the second (queries **L6**–**14**) having an $\mathcal{M}$ value less than 1. Each of the queries is executed with three different plans: $D$, a direct evaluation with a single wavefront with no threading; $T_1$ performs a threading split on predicate $p$; and $T_2$ threads on locatedIn+.

The relative running times for queries **L1**–**14** executed with plans $D$ (the baseline), $T_1$, and $T_2$ are presented in Fig. 18. As anticipated, the evaluation of queries in the first group is significantly (up to 75%) faster threaded than direct, and with $T_1$ being slightly faster than $T_2$. This can be attributed to that the length of the shared path $\mathcal{L}$ is shorter in $T_2$ due to a "later" threading split in the query expression. Also as anticipated, queries in the second group show bad results for $T_1$. Indeed, picking a predicate with $\mathcal{M} < 1$ for a threading split will generally be bad due to few shared paths. On the other hand, the results for $T_2$ are better than $D$ for 5 out of 9 queries in this second group. This is explained by the lensing effect, which is produced by concatenation $p/$:locatedIn+, while $\mathcal{M}(G, p) < 1$ and $\mathcal{M}(G, \text{:locatedIn+}) > 10$. Depending on whether $\mathcal{M}(G, p/\text{:locatedIn+})$ is greater than or less than 1, threading is either desirable or not, respectively.

### 5.4 Loop Caching

We benchmark the loop-caching optimization by executing a collection of queries of the simple template $Q_{(ab)+} = (x, (ab)+, y)$ ($\mathcal{Q}_2$) with two WGPs $P_{nc}$ and $P_{fc}$, which specify executions of $Q$ with no loop caching and with full loop caching, respectively.

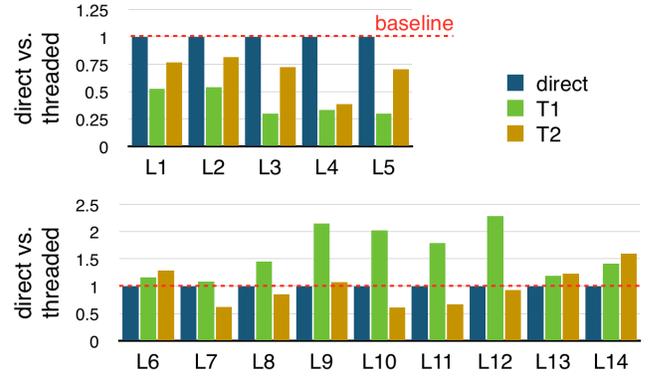

Figure 18: Threading over YAGO2s.

Values for $a$ and $b$ were chosen by iterative pruning of predicates appearing in the DBPedia dataset. First, we excluded predicates with very high (more than 25M) and low (less than 75K) cardinalities. Then, we ran query $Q_{abab} = (x, (abab), y)$ and recorded those $(a, b)$ predicate pairs for which the result of $Q_{abab}$ was not empty. DBPedia had 1171 such pairs, which indicates a high number of $(ab)+$ paths in this dataset. For each of these pairs, we ran the full closure query $Q_{(ab)+}$ to obtain its *expansion ratio*,

$$r_{\exp} = \frac{|Q_{(ab)+}|}{|Q_{ab}|} \qquad (\mathcal{E}_2)$$

where $|Q|$ denotes the cardinality of a query result.

Recall that both $P_{nc}$ and $P_{fc}$ initially evaluate $(ab)$ paths in the same way, while the rest of the closure $(ab)+$ is computed differently. Hence, in order to show the differences between these plans, we chose predicate pairs with $r_{\exp} \gg 1$, so that the computation of the rest of the closure constitutes the majority of the plan execution time. We identified 38 such queries by analyzing graph patterns in DBPedia.

We evaluated each one of these queries with $P_{nc}$ and $P_{fc}$ plans and recorded the *running time*, *edge walks* and *pruning statistics*. Due to widely varying absolute values for these statistics across queries, we present their relative percentage breakdowns in Fig. 17, as follows. Each query is represented by a two-colored bar, which shows the percentage breakdown of statistics values between $P_{nc}$ and $P_{fc}$ executions. In this way, we present edge walks (in the left chart) and running-time execution (in the right chart). We enumerate the queries from **D1** to **D38** according the ascending sorting of the percentage of edge walks performed in the $P_{nc}$ execution relative to the $P_{fc}$ execution. Hence, in query **D1**, $P_{nc}$ execution resulted in significantly fewer edge walks relative to $P_{fc}$ execution, with the opposite true for query **D38**. Finally, we perform a further breakdown, for each query, of the total number of edge walks into the number of tuples which were cached, were reduced against the cache, or were reduced against the delta. This breakdown is represented by different shades of the color associated with $P_{nc}$ or $P_{fc}$ executions, respectively.

Our first observation is that, in general, the loop caching optimization can significantly increase *or* decrease the total number of edge walks performed by the search. In our benchmark, loop caching resulted in fewer edge walks in 68% of the queries, with almost an order of magnitude reduction, in the best case. On the other hand, in 32% of the queries,

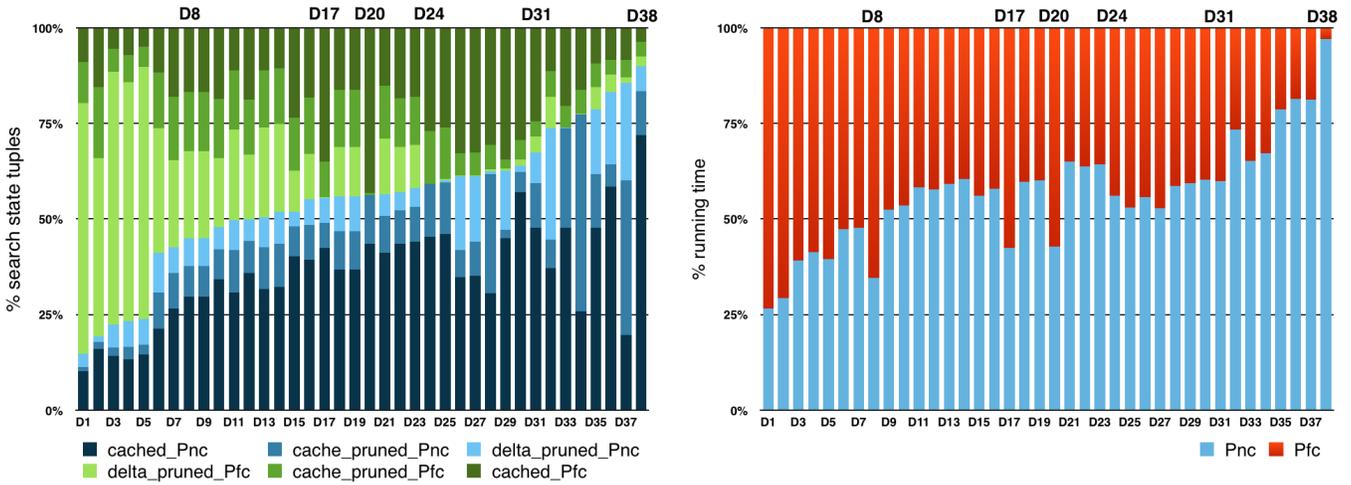

Figure 17: Edge walks vs. runtime in plans w/ & w/o loop caching in DBPedia.

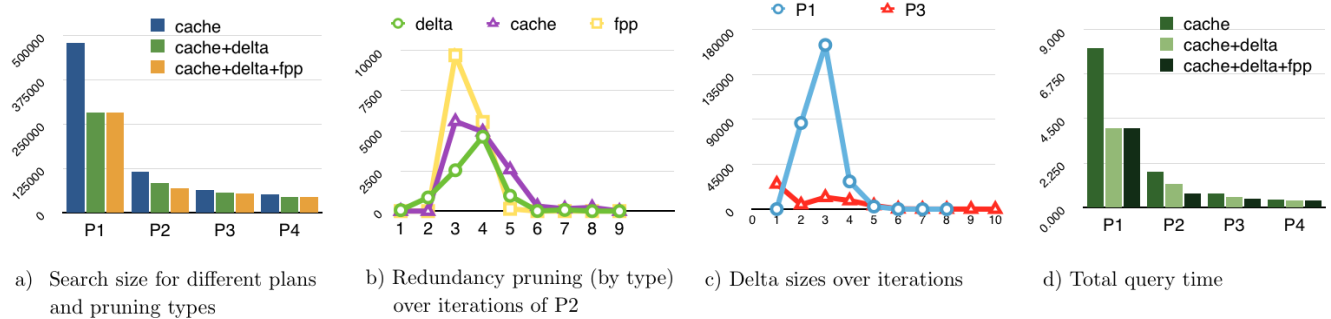

a) Search size for different plans and pruning types

b) Redundancy pruning (by type) over iterations of P2

c) Delta sizes over iterations

d) Total query time

Figure 19: Effect of plans on query evaluation.

loop caching resulted in more edge walks, with a more than 5X increase, in the worst case.

Our second observation is that the query running time is correlated to the total number of edge walks performed, but with some deviations. In queries with bad loop caching performance (**D1**-**D8**), the running time grows more slowly than the number of edge walks. This is due to that, in these queries, the majority of edge walks produced duplicate tuples, which were removed against the delta. Such removals are inexpensive, as discussed in §4.1. On the other hand, due to the lack of delta removals in edge walks, we observe an increase in the running time relative to the number of edge walks in **D17, D20, D24-31**, and **D33-34**. The running time for outliers **D8** and **D38** is affected by the cost of intermediate concatenations performed during the evaluation. Simple cost estimates (based on the product of relations) for cranks over iterations are presented in Fig. 15. In **D8**, $|C_{ab}| \gg |C_a|$ and $|C_{ab}| \gg |C_b|$, which slows down the concatenations in $P_{lc}$ when compared to $P_{nc}$. The opposite is observed in **D38**, yielding the advantage to $P_{fc}$ over $P_{nc}$.

Lastly, we study the effect of lensing by analyzing the degree of delta and cache pruning. Fig. 16 plots pruning over iterations for the queries which exhibit lensing: **D3** and **D38**. Query **D3** has $\mathcal{M}(G, a) = 10.58$ and $\mathcal{M}(G, b) = 0.33$, which suggests lensing with focal point on the concatenation $a/b$. As discussed in §4.4, this can significantly increase amount of pruning for loop caching, which is indeed what we observe. On the other hand, **D38** has $\mathcal{M}(G, a) = 0.07$ and $\mathcal{M}(G, b) = 5.34$, which suggests lensing with focal point on the concatenation $b/a$. This lensing benefits loop caching by decreasing the amount of pruning over iterations, which is what we observe.

## 5.5 Combined Optimizations

We illustrate the impact of combining WAVEGUIDE optimizations over the example query

?p :marriedTo/:diedIn/:locatedIn+/:dealsWith+ USA ($\mathcal{Q}_4$)

over the YAGO2s dataset. We instantiate p as follows.
P$_1$: single wavefront USA → ?p.
P$_2$: single wavefront ?p → USA.
P$_3$: two wavefronts
　　　　?p → :locatedIn+/:dealsWith ← USA.
P$_4$: P$_2$ but with a threaded sub-path
　　　　:locatedIn+/:dealsWith+ USA.

Fig. 19a shows the effect of wavefront choice on search cardinality. Note the order of magnitude difference between the best, P$_4$, versus the worst, P$_1$. The three types of redundancy pruning—cache, delta, and fpp—are illustrated for each plan. Fig. 19b plots search size across iterations for P$_2$ with pruning; over 40% of tuples are pruned! Fig. 19c plots delta sizes over iterations for P$_1$ and P$_3$. Note how the selective search of P$_3$ is better behaved than the rapid expansion of P$_1$. In Fig. 19d, the total execution time for each plan is presented. This demonstrates the significant improvement in performance achievable by careful design of the WGP.

## 6. NEXT STEPS & CONCLUSIONS

Waveguide plans model a rich space of plans for path queries which encompass powerful optimization techniques. Next steps in this endeavor are as follows.

1. *Benchmark* Waveguide against current, prevalent SPARQL engines that support property path queries (e.g., Jena TDB, Virtuoso, and AllegroGraph).
2. *Build* a full-fledged *cost-based query optimizer* for SPARQL 1.1 for property paths (RPQs).
   (a) *Define* "WGP" systematically to define formally the *space* of WGPs for a given query.
   (b) *Devise* a concrete *cost model* for WGPs.
   (c) *Determine* an array of *statistics* (e.g., 1-gram and 2-gram label frequencies) that can be computed efficiently offline that can be used in conjunction with the cost model.
   (d) *Design* an enumeration algorithm to walk dynamically the space of WGPs to find the WGP with least estimated cost.
3. *Extend* the *query optimizer* to handle queries queries with multiple property-paths (equivalent to *conjunctive* regular path queries).

Just as new data models necessitate new query languages, these new query languages necessitate new approaches if we are to evaluate their queries efficiently and effectively. The rise of graph databases has necessitated new, powerful query languages so that we can make use of them. But we are only beginning to uderstand how we can deal effectively with these types of queries.

In this work, we have devised a rich domain of evaluation plans for property-path type queries in SPARQL, and have shown it extends significantly over the state of the art. We have demonstrated that choice of plan can make orders of magnitude difference in performance. We have illustrated the cost factors behind these plans's performance and the types of optimizations that can be achieved. We have shown which plans are effective depends on the underlying graph database, which means a cost-based means of choosing plans is required. The rise of graph data is well underway. And as we learned in the past to do the "impossible" for relational data, for semi-structured, for unstructured search, we too will meet this challenge.